# FEMORAL FRACTURE LOAD AND DAMAGE LOCALISATION PATTERN PREDICTION BASED ON A QUASI-BRITTLE LAW: LINEAR AND NON-LINEAR FE MESHING

**Zahira Nakhli [1], Fafa Ben Hatira[1],Martine Pithioux[2,3], Patrick Chabrand[2,3]**

## Khemais Saanouni[4]

[1] *Laboratoire de Recherche Matériaux Mesures et Application (MMA), LR11ES25,University of Carthage, National Institute of Sciences and Technology (INSAT), Tunis, Tunisia.*
[2]*Aix Marseille University, UMR CNRS 7287, (ISM) Institute of Movement Sciences, Marseille, France.*
[3]*APHM, Ste-Marguerite Hospital, Inst for Locomotion, Dep of orthopaedics & Traumatology, Marseille, France*
[4] Laboratoire des Systèmes Mécaniques et d'Ingénierie Simultanée Institut Charles Delaunay, UMR CNRS 6281, UTT, TROYES, France

## Abstract

Finite element analysis is one of the most used tool for studying femoral neck fracture. Nerveless, consensus concerning either the choice of material characteristics, damage law and /or geometric models (linear on nonlinear) still remains unreached.

In this work, we propose a numerical quasi-brittle damage model to describe the behavior of the proximal femur associated with two methods to evaluate the Young modulus. 8 proximal femur finite elements models were constructed from CT scan data (4 donors, 3 men; 1 woman). The results obtained from the numerical computations showed a good agreement between the numerical curves (load – displacement) and the experimental ones. The computed fracture loads were very close to the experimental ones ($R^2$=0.825, Relative error =6.49%). The damage patterns were similar to those observed during the failure during sideway fall experimental simulation. Finally, a comparative study based on 32 simulations, using a linear and nonlinear mesh has led to the conclusion that the results are improved when a nonlinear mesh is used.

In summary, the numerical quasi-brittle model presented in this work showed its efficiency to find the experimental values during the simulation of the side fall.

## Introduction

The osteoporosis disease, which is defined as a decrease in bone strength, can be estimated by bone mineral density (BMD) measuring [1]. This pathology causes fractures in different bone structure and is classified as the most important ones affecting the femoral neck [2,3,4]. It usually occurs without apparent symptoms until the provocation of the fracture. Fracture prevention if this pathology based on diagnosis can delay surgical procedures. Finite element (FE) modeling can be a reliable tool to better screen up the different factors related to bone fractures and give surgeons more reliable criteria on fracture risk factor. Indeed, numerical modeling based on Finite element method has appeared in the 1950s and helped engineering to deal with different problems of structural mechanics. Some specific models were developed in mechanical to predict human proximal femur fracture and to assess the pressure distribution under physiologic loading in bone structures.

Most of proposed models were focusing on the prediction of the ultimate force at fracture as well as the fracture pattern by using different mechanical approaches. These studies were based on linear and non-linear isotropic and /or anistropic FE models [5,6,7,8,9,10,11,12,13,14]

The various works mentioned above have tried to give a unique answer to the solution of modeling the behavior of bone structures and more precisely in the context of this work to the problem of fracture of these structures. No consensus has yet been reached, but each scientific work carried out can help to move towards a construction of an efficient prediction. Most of the last studies were carried out with nonlinear FE modeling for a better efficiency of the proximal femur fracture. In order to propose a new efficient numerical tool, inexpensive (from computing side of view) and of course close to experimental, a method for estimating the proximal femur fracture based on a non linear FE model is presented in this study. The Continuum Damage Mechanics CDM framework is chosen to develop the isotropic quasi-brittle fracture law with two elasticity properties (homogenous and non homogenous). The model is implemented into a user routine VUMAT in the finite element software (Abaqus) and applied in linear and non linear meshing model. The Finite element simulations were carried out using the explicit dynamic algorithm. Numerical computations for four osteoporotic femurs (right and left, height specimens) were compared with success to experimental fracture data (values and curves) with linear and non linear meshing.

**Méthode**

### CT Scan

Eight femurs (right and left) coming from four donors (3men, 1 women), were scanned individually with high resolution by using alight speed VCT scanner from GE Medical Systems available in Medical imagery service of La Timone University hospital, (Marseille). The resolution system used provides a three-dimensional map of the bone mineral density through the studied bone structures. [9,15,16]

The different steps taken to apply the protocol required to create the finite element model from CT data are described in Figure 1.

The first step was the reconstruction of the 3D geometry of each femur from the Xrayscanner images based on the generation of the voxel element using the research software Mimics 17.0. Densities described by grey value level were assigned to each voxel element [17].

The second step was the femur volume 3D mesh generating with tetrahedral elements by the research software 3Matic 9.0.0.231. In order to assign the material parameters, the volume mesh was finally imported a second time in Mimics (step 4). Thereby, a 3D model specific to each patient respecting his anatomy and possessing material properties related to the quality of his bone was created and imported to the Abaqus/CAE software (step 5). More details of this protocol can be found a previous article[18].

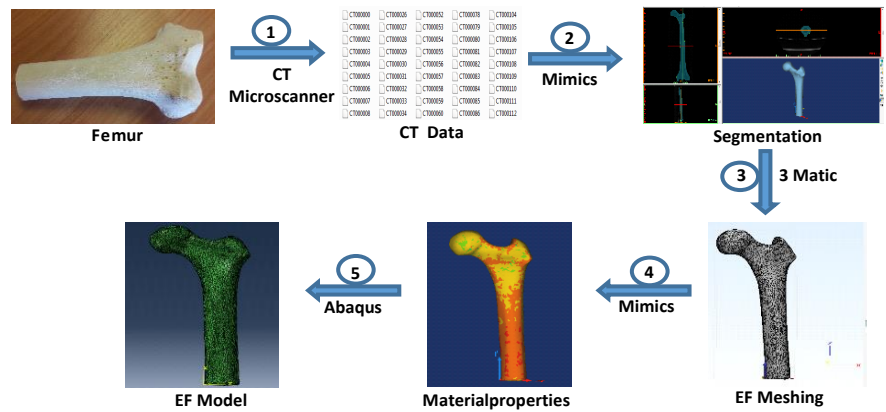

**Figure 1.** The protocol established to create a 3D FE Model from computed tomography data using the element-by-element material properties using the Hounsfield unit of CT data.

**Experimental Mechanical Compression test**

To reproduce a simulation of a sideways fall on the greater trochanter, each proximal femur was loaded to failure in the INSTRON 5566 machine. Specimens were fixed in resin (Epoxy Axon F23) at 15.12° internal rotation. The femoral shaft was oriented at 10° adduction in the apparatus (Figure 2 left). For this Specimen in Figure 2 Right, the neck forms an angle with the shaft in about 125.08° degrees, which is called diaphysealangle, in this case it is a Normal diaphysealangle (between 120° and 137°). The load was applied to the greater trochanter through a pad, which simulated a soft tissue cover, and the femoral head was molded with resin to ensure force distribution over a greater surface area [19]. The figure 2 shows the mechanical test conditions of thesideway fall simulation.

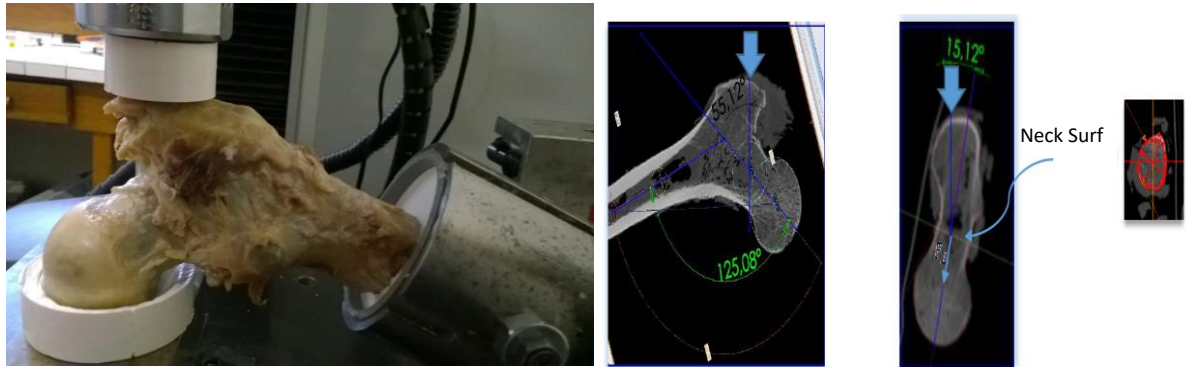

**Figure 2-** The mechanical compression test conditions
(left) :CT scans before fracture (Right) : Normal diaphysealangle125.08°, Femur was fixed in resin (Epoxy Axon F23) at 15.12° internal rotation

The results of the different conducted experiences are reported in Table 1. It gives the details of the obtained ultimate failure load and the bone mineral density (BMD) for all femurs, which are composed of three osteoporotic femurs and one healthy. The mean value of the BMD was found to be 0.7017 g/cm2.

**Table 1:** Failure values for the eight proximal Femur

| Specimen | femur | BMD(g/cm2) | Load (N) |
|----------|-------|------------|----------|
| Sp1L     | Osteoporotic | 0.651 | |
| Sp1R     | Osteoporotic | 0.722 | 1524.42 |
|          |              |       | 2318.69 |
| Sp2L     | Osteoporotic | 0.615 | |
|          |              |       | 973.00 |
| Sp2R     | Osteoporotic | 0.508 | 743.00 |
| Sp3 L    | Osteoporotic | 0.714 | |
|          |              |       | 1477.17 |
| SP3R     | Osteoporotic | 0.701 | 1293.23 |
| Sp4L     | healthy | 0.842 | |
|          |         |       | 1493.98 |
| Sp4R     | healthy | 0.861 | 1113.92 |
| Mean value | | 0.7017 | |

From experimental data, eight load-displacement curves are plotted in Figure 3. The obtained curves for right and left femurs taken from the same donor showed different tendency with different fracture load magnitude. Exception is however reported for the curves of Specimen 2 where similarity is noticed for the left and the right femurs.

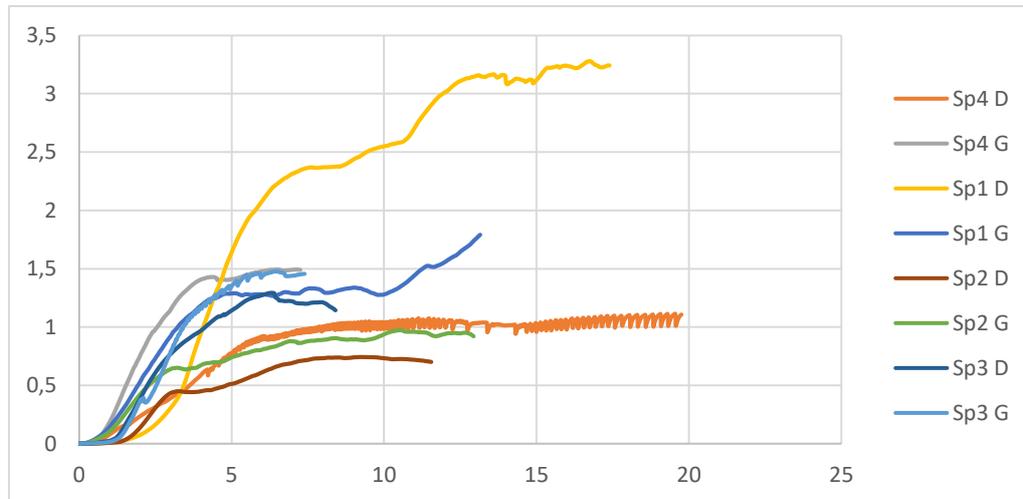

**Figure 3 -** Experimental load vs. displacement for the eight specimens

## Quasi Brittle LAW

The model is based on an isotropic behavior law coupled to a quasi-brittle damage law in order to describe the progressive initiation and propagation of cracks within human proximal femur under quasi-static load.

In this work, the damage behavior law describing a quasi-brittle behavior is proposed using an isotropic Continuum Damage Mechanics (CDM) based on Marigo[20] modeling of the damage for brittle and quasi-brittle material.

- <u>Constitutive framework: A quasi brittle damage model</u>

The approach of irreversible thermodynamics with internal variables(Chaboche1988;Germain1973;Krajcinovic1989;;Kachanov1986; Lemaitre1987, Saanouni et al. 1996, Saanouni 2012) is chosen to present a coupled damage elastic model to describe the initiation and the accumulation of the damage in bone structure, more precisely the femur.

The new energy based model is described throughout state variables (external and internal).Thestate variables describing the constitutive equations are represented by theexternal and the observable state variables, namely the elastic strain tensor $\underline{\varepsilon}^e$ andand the Cauchy stress tensor$\underline{\sigma}$. For the sake of simplicity, damage is supposed isotropic described by a couple of scalar internal variables (D,$Y$) where $Y$ is the damage force associated to the damage variable D.

The effective variables $\underline{\tilde{\varepsilon}}^e$ and $\underline{\tilde{\sigma}}$ including the damage effect, which are defined in the framework of the elastic strain equivalence assumption, are presented hereafter. The expression of the stored elastic energy density is given by:

$$\rho\psi\left(\underline{\varepsilon}^e, D\right) = \frac{1}{2}(1 - D)\underline{\varepsilon}^e : \underline{\underline{A}} : \underline{\varepsilon}^e + \widehat{\psi}(D) \quad (5)$$

According to the Marigo hypothesis: $\widehat{\psi}(D) = 0$ dire ce que c'est $\widehat{\psi}$

Where $\underline{\underline{A}}$ is thesymmetric fourth-rank tensor of elastic properties of the virgin (not affected by damage) material, which in the isotropic case can be written in terms of the well-known Lame's constants$\lambda$and $\mu$according to:

$$\underline{\underline{A}} = \lambda\,\underline{1} \otimes \underline{1} + 2\mu\,\underline{\underline{1}}$$

$$\mu = \frac{E}{(1 - 2\upsilon)}; \quad \lambda = \frac{E\upsilon}{(1 + \upsilon)(1 - 2\upsilon)}$$

Where $\underline{1}$ is the second-rank identity (Krönecker) tensor while $\underline{\underline{1}}$ is a fourth-rank unit tensor.

According to the theory of Marigo, the global energy depends only on the two state variables namely the elastic strain tensor and the damage.

The state laws$\underline{\sigma}$and Y,are classically derived from the state potential are obtained from the freeenergy by:

$$\underline{\sigma} = \rho\frac{\partial\psi}{\partial\underline{\varepsilon}^e} = (1 - D)\underline{\underline{A}} : \underline{\varepsilon}^e \quad (6.1)$$

$$\underline{\underline{A}} = \rho\frac{\partial^2\psi}{\partial^2\underline{\varepsilon}^e} > 0 \qquad\qquad (6.2)$$

$$\underline{\sigma} = (1 - D)\left(\lambda\,\underline{1} \otimes \underline{1} + 2\mu\,\underline{\underline{1}}\right) : \underline{\varepsilon}^e \quad (6.3)$$

$$Y = -\rho\frac{d\psi}{dD} = \frac{1}{2}\underline{\varepsilon}^e : \underline{\underline{A}} : \underline{\varepsilon}^e \quad (7)$$

The damage criterion(or damage yield function) is described by Y:

$$f(Y, D) = Y - \frac{1}{2}Y_0 - mD^{\frac{1}{s}} = 0 \qquad (8)$$

where $Y_0$,s and m are material parameters. The parameters s and m are related to the damage "hardening" of the material. It is here to be noticed that the damage yield function Eq.(8) can describe the initiation of micro-cracks starting from undamaged state (D=0).

In the present model, the dissipation potential $\boldsymbol{\varphi}$ is reduced to the yield function f according to the associative theory:

$$\varphi = f = Y - \frac{1}{2}Y_o - mD^{\frac{1}{s}} = 0 \quad (9)$$

The damage evolution equation derived from the dissipation potential is:

$$\dot{\varphi} = 0 \Leftrightarrow \frac{\partial \varphi}{\partial Y} \dot{Y} + \frac{\partial \varphi}{\partial D} \dot{D} = 0 \quad (10)$$

For this approach, the coupling between damage and elasticity is completed with the following damage evolution law.

$$\dot{D} = \frac{s}{m} \frac{\dot{Y}}{D^{\frac{1-s}{s}}} (11)$$

With:

$$\dot{Y} = \underline{\varepsilon}^e : \underline{\underline{A}} : \dot{\underline{\varepsilon}}^e \quad (12)$$

According to Eq. (6.3), when damage increases by Eq. (11), then the stress tensor decreases due to the decrease of the Lame's constants (i.e. the Young's Modulus).

Solving the non linear problem described by Eqs (6)-(12) in order to determine the unknowns of the problem is performed through an approximation of these variables in total time interval $I_t = [t_0, t_f] = \bigcup_{n=0}^{Nt} [t_n, t_{n+1} = t_n + \Delta t]$, $\Delta t$ being the increment between two successive time steps. This approximation is done for every integration point related to every finite element.

Thus knowing the initial variables at $t_n$, the discretized problem is solved giving the final solution at the final time $t_{n+1}$. The discretization leads to the following expressions of the problem variables at $t_{n+1} = t_n + \Delta t$, the end of the step time:

$$\underline{\sigma}_{n+1} = (1 - D_{n+1}) \left( \lambda \mathrm{tr} \, \underline{\varepsilon}^e_{n+1} . I + 2\mu \underline{\varepsilon}^e_{n+1} \right) (13)$$

$$\underline{\varepsilon}^e_{n+1} = \underline{\varepsilon}^e_n + \Delta \underline{\varepsilon}^e_n (14)$$

$$Y_{n+1} = \frac{1}{2} \underline{\varepsilon}^e_{n+1} : \underline{\underline{A}} : \underline{\varepsilon}^e_{n+1} (15)$$

$$f_{n+1} = Y_{n+1} - \frac{1}{2} Y_0 - m D_{n+1}^{\frac{1}{s}} = 0 \qquad (16)$$

From this last equation, the "admissible" value of the damage variable is deduced as:

$$D_{n+1} = \langle \frac{Y_{n+1} - \frac{1}{2} Y_0}{m} \rangle^s \qquad (17)$$

In this isotropic damage model, some remarks can be made:

- If the scalar variable describing damage is zero (D=0.0), then the material state is described by the classical isotropic elastic model.
  If the fracture condition of the critical value (D=1.0) is reached, then the material point is declared as fully damaged and the following value is assigned to D in that point D=0.999.

**The proposed Algorithm**

As mentioned earlier, one of the goals of this work is to propose a reliable quasi brittle damage model of proximal Femur fracture based on the finite element model ( FEM).The proposed algorithm followed to implement the model is summarizedin figure 4.

The first stepconsists of the global modeldefinition: geometry, load conditions and initial bone density distribution. The Second Step is concerned with the determination of Young's modulus, Poisson's ratio and the density. These material properties are obtained by two methods. The first one, the homogeneous material distribution calculate through the research software Mimics 17.0, based on the Hounsfield Units 'HU' or Units Gray values on the scanned images. The second one, the heterogeneous material distribution, assigns to each element a distinct mechanical property. The third step is related to the calculation of the displacement, by solving the linear variational equation of the displacement field. During step 4, the evaluation of the strain, the stress and the damage at each discrete location is performed and it is based on the finite element method. Thereafter, an update of the stress (step 5) and damage (step 6) values are applied. The model being implemented into the Abaqus/Standard code using the subroutine UMAT, a check for convergence is executed. The final result is obtained when the convergence criterion is satisfied; otherwise, the iterative process continues from Step 2.An illustration of the algorithm used is described in Figure4.

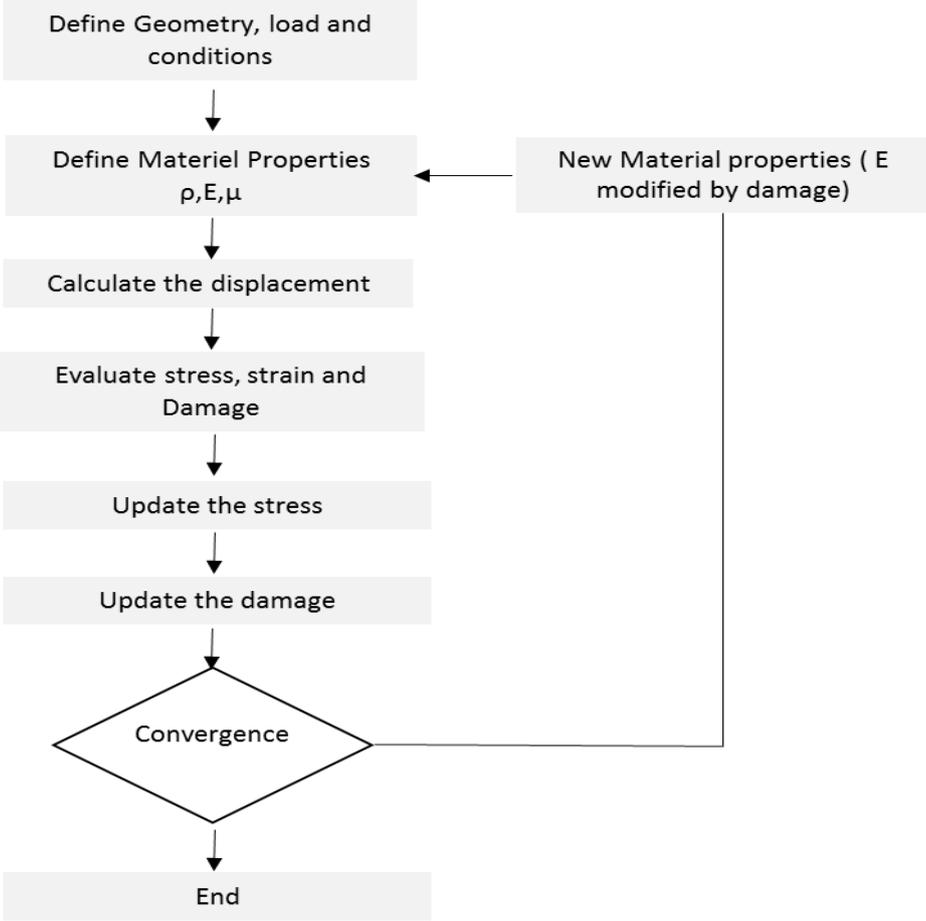

**Figure 4**- Schematic representation of the Bone Damage algorithm proposed.

In the present work and for the sake of comparison, bone was modeled as an isotropic material with two mechanical properties (Young modulus E (MPa)) estimated through two different methods. These techniques were used in previous works. Indeed, most of these studies, adopted a non homogenous elastic modulus related the density. As examples of these works we can recallmorgan E.F et al. [21], Keyak J.H. [22], Ariza O. [23], Pithioux et al. [24] and Haider et al.[7]. However, some researchers adopted the method based on the assumption that the elastic modulus is homogenous and a function of volume fraction (BV/TV) . These studies can found in the references (Hambli R. et al. [25] and Varga P. et al. [13]). We will recall here after these two methods of E determination.

Method 1 : Non homogenous Young Modulus

The first method using the following expression gives a varying Young modulus (E) related to the bone apparent density ρ such as defined by Kaneko et al. [26],

$$E_1 = 2000 \, \rho^{1.89} \qquad (1)$$

ρ: apparent density (g/cm$^3$)

This method is based on a phenomenological law and allows to assign to each element a distinct mechanical property using a direct correlation between apparent density and Young Modulus. In the end, a heterogeneous material distribution was obtained.

Method 2 : Homogenous Young Modulus

For method 2, homogenization techniques were considered, allowing to obtain a homogeneous material distribution. The following relationship is proposed by Hernandez et al. [27]:

$$E_2 = 84370 \left(\frac{BV}{TV}\right)^{2.58} \qquad (2)$$

BV/TV:bone volume/total volume fraction;

Thirty-two numerical computations based on eight Femurs reconstructions are validated through a comparison of the experimental crack localization and of the estimated failure loads.

The material properties E and ρ are summarized in Table 2. Poisson ratio is set at 0.3 based on the work of [28,29,30].

**Table 2.** Material Properties for height specimens

| Femur | | Strength Load (N) | Density | Young Modulus (Mpa) Method 1 | Young Modulus(Mpa) Method 2 | Poisson ratio |
|---|---|---|---|---|---|---|
| Sp1 | G | 1524.42 | 0.28 – 2.45 | 121 – 14475 | | |
| | D | 2318.69 | 0.25 – 2.46 | 103 - 14552 | | |
| Sp2 | G | 973 | 0.47 – 2.44 | 384– 14293 | | |
| | D | 743 | 0.35 – 2.42 | 199 - 13991 | 3777 | 0.3 |

| Sp3 | G | 1477.17 | 0.41-2.41 | 292-13905 | | |
|-----|---|---------|-----------|-----------|--|--|
| | D | 1293.23 | 0.32-2,46 | 161-14469 | | |
| Sp4 | G | 1493.98 | 0.34-2.458 | 185-14526 | | |
| | D | 1113.92 | 0.31-2.45 | 151-14405 | | |

## Simulations

- **Boundary and Loading conditions**

The numerical validation is conducted with the boundary condition and load case representing the experimental conditions (Figure 5). The load was applied on femoral Greater trochanter reproducing the sideway fall case, whereas the femoral head and the lower surface were constrained. During the conducted computations, the stiffness of elements degrades gradually as damage increase, and the crack is modeled as the region of elements whose stiffness has been reduced to near zero.

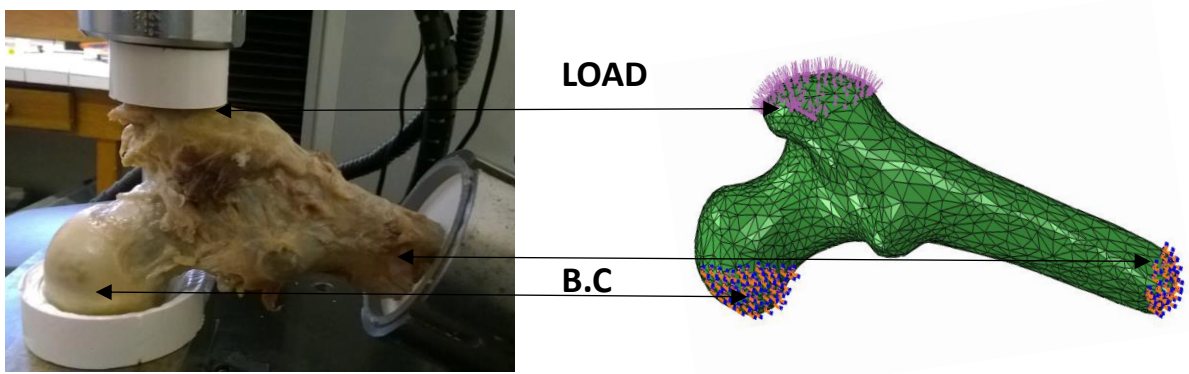

**Figure 5 -**The mechanical compression test conditions (Left) and BCs applied to the FE model(Right)

## Results

The general purpose of this work is to compare the prediction of the damage localization as well as of the ultimate fracture load for different specimens tested experimentally for two meshing model, linear mesh with linear tetrahedral elements (C3D4), and nonlinearmesh with quadratic tetrahedral elements (C3D10) (six degrees of freedom per node which are the three displacements and the three rotations). The ultimate strength load value obtained experimentally was applied for the two methods which are based on two different approaches to estimate the Young modulus as previously presented. The analysis details the fracture load and the localized damaged zones dependency on the young modulus estimation as well as the linearity or not of the meshing.

The 32 correlations between the experimental and FE computed fracture loads for the four studied cases are exposed in Figure 6. In summary and as it appears in Figure 6 (D), the numerical computations based on the use of an homogeneous material distribution with a nonlinear mesh, present a good agreement with the experimental data (fracture load magnitude) with the best correlation ( $R^2$= 0.825). For the other cases (Figure 6 (A),(B) and(C)) , correlations are weak and lower than 0.356 .

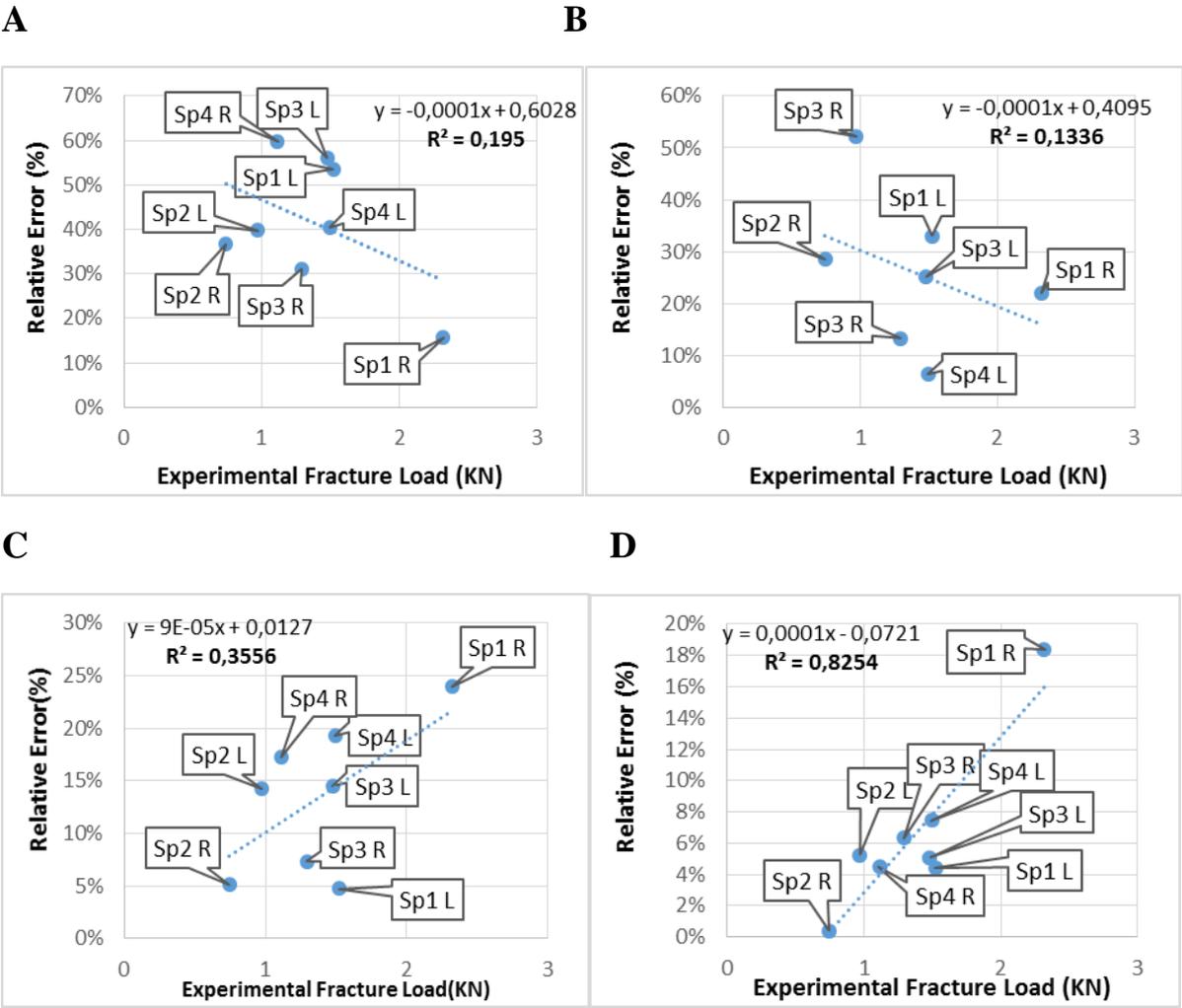

**Figure 6-** Numerical fracture loads (KN) and the relative error based on a comparison with the experimental data for the eight cases (right and left femurs) computed with $E_1$ and $E_2$: (A) Method1-Linear Mesh; (B) Method1-NonLinear Mesh; (C) Method2-Linear Mesh; (D) Method2-NonLinear Mesh

In Table 3 the relative errors between the experimental and the computed fracture loads are reported. The best results is obtained for the case where we use Young modulus E related to the microarchitecture parameter (method 2) for the non-Linear mesh. The fracture load error average was found to be 6.49 %.

**Table 3.** Relative Error´s summary

|  | Linear Model | | Non Linear Model | |
|---|---|---|---|---|
|  | Relative Error (%) | Average (%) | Relative Error (%) | Average (%) |
| Method 1 | 15.6 - 59.7 | 41.63 | 6.5 - 52.2 | 23.95 |
| Method 2 | 4.7 - 24 | 13.31 | 0.4 - 18.4 | 6.49 |

The propagations of the cracks and the distribution of the quasi-brittle damage of the eight femurs are plotted in Figure7

The results of the numerical computations gave two different crack localizations based on the choice of the elastic property. Indeed, the FE simulations performed with the method 2,showed a femoral neck (transcervical) fracture, the crack is initiated locally at the superior surface of femoral neck. Then the crack continues to grow, resulting a separation of the proximal femur. However, in this case the damage surface corresponded to the fracture surface observed in the experience, differently from the first method, where fracture occurred in the Greater trochanter. The same observations are obtained for the two femurs (right and the left) for the all studied femurs.

The crack localizations for the two models linear and nonlinear are quasi similar, except that in the nonlinear modeling case, the entire femur is affected by the damage as shown in Figure 7 for the SP4 R and SP4 L. These observations are very interesting since as it has been related in previous studies, nonlinear meshing can be computationally expensive (cpu).

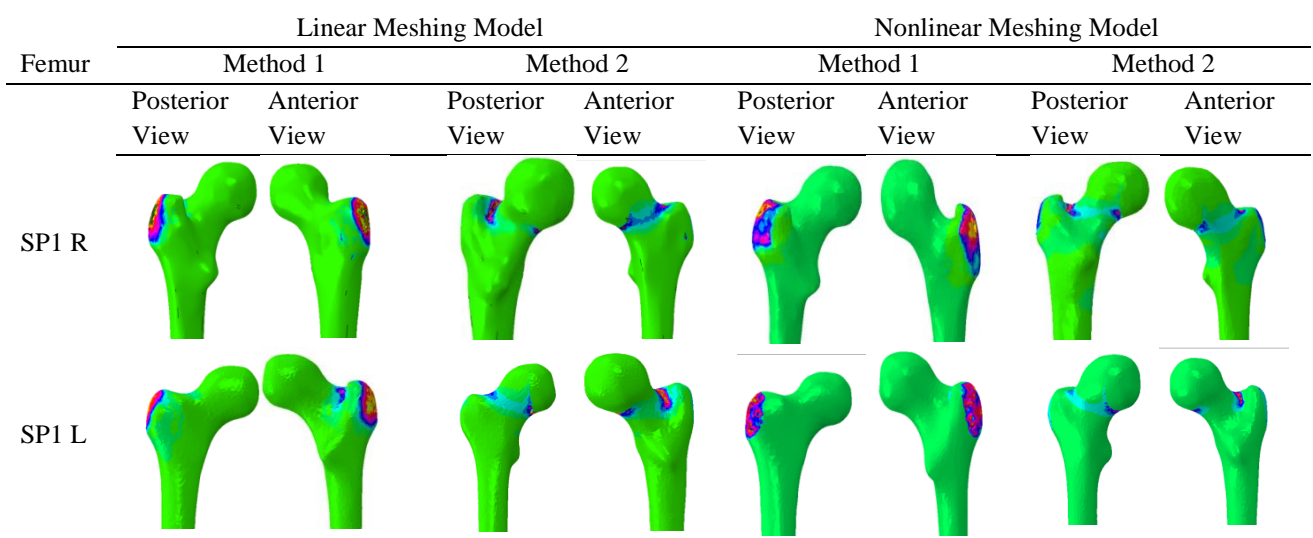

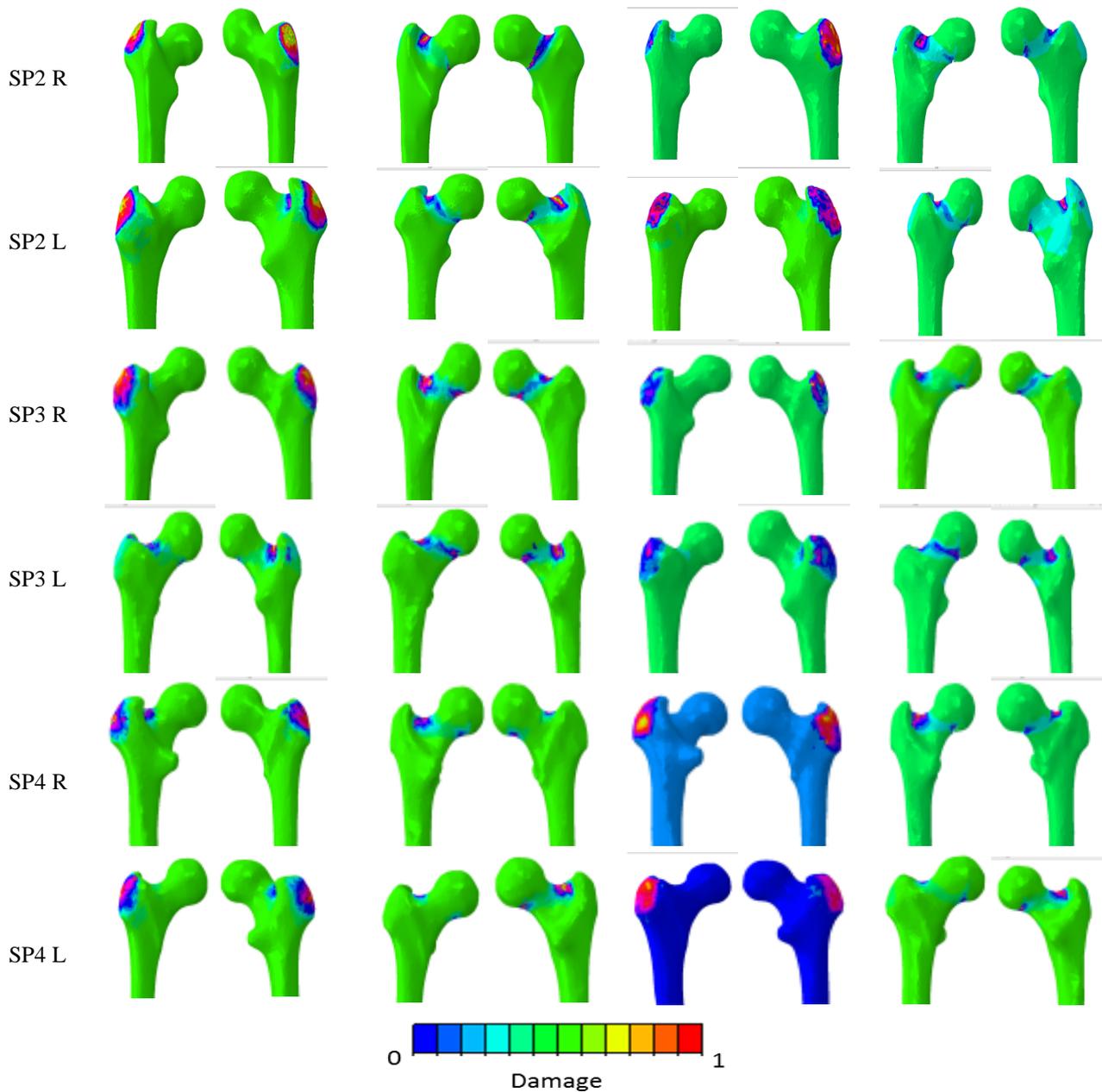

**Figure 7** - Predicted fracture pattern from different view and quasi-brittle damage distribution.

Special attention will now be paid to one of the specimens presented in the previous overall results. The final goal is to better underlined the quantitative and qualitative results obtained for the eight specimens. The specimen chosen is the one named SP3R. It is a representative sample of all the studied specimens. We begin by a comparison between the experimental and numerical behavior curves (with linear and non linear meshing) obtained during the simulation of the sideways fall.

This comparison which is illustrated in Figure (8) shows a good simulation between experimental and numer    **LinearMesh**    es, the nonlinear case shows the best agreement with experimental curve, as well as a sharp drop in force during failure.

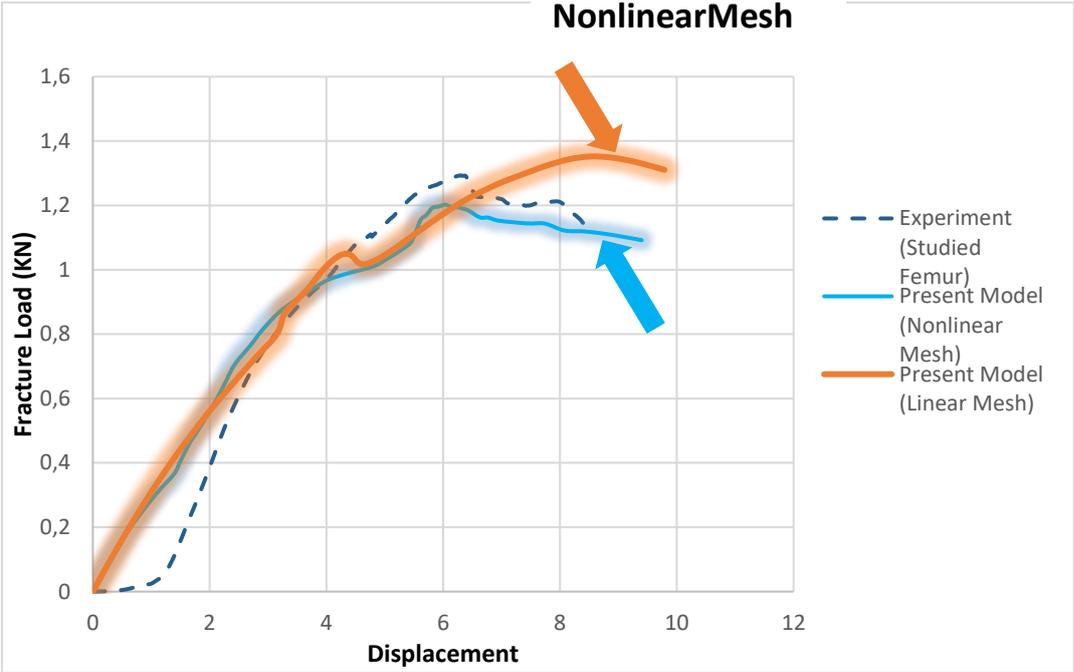

**Figure 8** -Predicted and experimental force-displacement curves of the present FE model Sp 3R for the Method 2, Linear and nonlinear mesh

Figure 9 shows a validation between the numerical for both cases of meshing /linear and nonlinear) and experimental results of the fracture pattern, and clearly demonstrate that the fracture line is located in the neck region

Also, Figures9a and 9b show that regardless of the choice of type of meshing (linear or nonlinear), we obtain fracture localizations similar to the ones obtained experimentally (Figure 9c).

In conclusion, this validation of load fracture and localization proves the performance of the adopted numerical method, which is formulated in the CDM framework. It also clearly demonstrated that the result is affected by the choice of the type of mesh (linear or nonlinear) whereas the damage pattern does not depend on this parameter.

| | Linear MeshModel | Nonlinear Mesh Model | Experimental |
|---|---|---|---|
| Femur | AnteriorView | AnteriorView | AnteriorView |
| SP3 R | 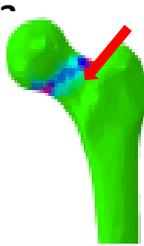 | 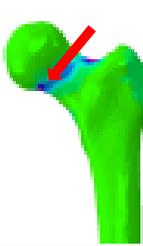 | 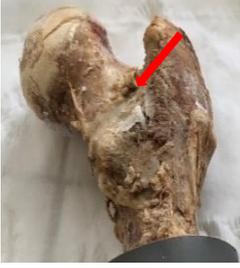 |
| SP3 L | 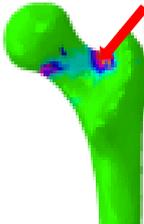 | 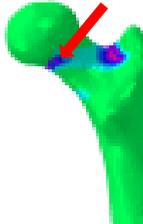 | 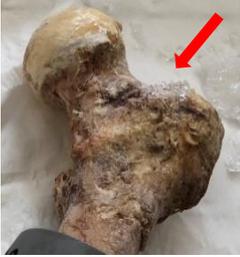 |

**Figure 9**-Qualitative evaluation of the FE based fracture pattern prediction, for linear mesh (a) and nonlinear mesh (b) by comparing with experimental compression-test photos (c) showing the anterior View for Specimen 3 Right and Left adopting method 2.

## Discussion

The aim of the present study was to implement a comparative study based on 32 simulations, using a linear and nonlinear mesh to show that the results are improved when a nonlinear mesh is used. As a first remark, we can say that the predicted force-displacement curve shows the same trend as the one observed experimentally. Regarding the relative error, the average error is about 6.49% with a very good fracture pattern predictions for all specimens, compared to previous works presetend by Haider I.T and al[7]. The average percentage errors of predicted fracture load was about 9.6% and peak error of only 14%. However, they have recalled that the average errors found in previous studies was less than the previously published studies which are from 10% to 20% [7].

In general, we found statistically moderate correlations between the experimentally and computationally results using a homogeneous young modulus (method2) and using nonlinear meshing(R²=0.825). However, no correlation was found between experimental and FE model for the heterogeneous young modulus distribution using the linear meshing and for two meshing model of the heterogeneous method. Regarding the localization issue, we found that in general, the experimental bone failure locations agreed with the locations of the FE for the method 2. Referring to the Garden Classification [31], different fracture can be observed experimentally and the numerically, depending on the femoral geometry, the material properties and the boundary conditions. In this work, the fracture patterns correspond to a Transervical neck fracture with stage II (Complete fracture with minimal or no displacement from anatomically normal position) of the Garden classification. To the best of the author's knowledge, this is the first time a comparison of linear and nonlinear meshing was established for the prediction of femoral fracture. Although we find the results encouraging, one limitation of this study is related to the mesh sensitivity, which should adopt an effective technique to better predict fracture load, fracture pattern, and fracture initiation independently of mesh density. Overall, the FE model precision was demonstrated by comparing the simulation results to the experimental results for each specimens.

**Conclusion**

The purpose of this work was to develop and validate a simple FE model based on continuum damage mechanics in order to simulate the complete force–displacement curve of femur failure. Femoral fracture load was predicted using a quasi brittle-damage FE model for four studied cases, combining homogeneous and heterogeneous material distribution with a linear and nonlinear mesh. The obtained results show a strong linear relationship between FE predicted and experimentally measured fracture load ($R^2 = 0.825$) in the case combining homogenous material distribution with non linear mesh. Furthermore, all eight cadaveric specimens present a similar failure locations between the experimental and the FE simulation, when the method 2 is adopting.

The presented FE model shows strong correlations between experimental and numerical values however is spurious mesh sensitivity. The size of the damaged region corresponds to the size of the mesh used to solve the problem. As the mesh is refined, the size of the damaged region

Despite limitations of our study, cited above, the relatively low average error in the fourth case suggests that this FE methodology may be useful in helping the surgeon choose a patient-specific treatment, and allowing them to make the right decision before the surgery by evaluating the risk factor from the fracture pattern.

**Acknowledgements**

The contribution of Saint Marguerite hospital radiology team specially P. Champsaur, T. Lecorroller, and D. Guenou is gratefully acknowledged.